\documentclass{elsart}
\usepackage{epsf}
\usepackage[psamsfonts]{amssymb}
\usepackage{amsmath}
\usepackage[dvips]{graphicx}
\usepackage[english]{babel}

\begin{document}

\begin{frontmatter}
\title{Fission and Nuclear Liquid-Gas Phase Transition }

\author{E.A.Cherepanov and V.A.Karnaukhov}
\address{Joint Institute for Nuclear Research,
141980, Dubna, Russia}

\begin{abstract}
The temperature dependence of the liquid-drop fission
barrier is considered, the critical temperature for the liquid-gas phase
transition in nuclear matter being a parameter. Experimental and calculated
data on the fission probability are compared for highly excited $^{188}$Os.
The calculations have been made in the framework of the statistical model.
It is concluded that the critical temperature for the nuclear liquid--gas
phase transition is higher than 16 MeV.
\end{abstract}


\end{frontmatter}

\section{INTRODUCTION}

The critical temperature for the liquid-gas phase transition is a crucial
characteristic related to the nuclear equation of state. There are many
calculations of $T_{c}$ for finite nuclei. In \cite{Sauer76, Jaqaman83,Shou96,Goodman84,Taras04},
it is done by using a
Skyrme effective interaction and the thermal Hartree-Fock theory. The values
of $T_{c}$ were found to be in the range 10-20 MeV depending upon the chosen
interaction parameters and the details of the model. In Ref. \cite{Randrup91,Lima92} the
thermostatic properties of nuclei are considered employing the
semi-classical nuclear model, based on the Seyler-Blanchard interaction. The
value of critical temperature is estimated to be $T_{c}$=16,66 MeV.

As the temperature of a nucleus increases, the surface tension decreases and
then \textit{vanishes }at $T_{c}$. For temperatures below critical, two distinct nuclear phases
coexist - liquid and gas. Beyond $T_{c}$ there is not two phase equilibrium,
only nuclear vapor exists.

The main source of the experimental information for $T_{c}$ is the yield of
intermediate mass fragments. In some statistical models of nuclear
multi-fragmentation the shape of the IMF charge distribution, $Y(Z)$, is
sensitive to the ratio $T$/$T_{c}$. It was noted in the earlier papers that the
fragment charge distribution is well described by the power law, $Y(Z) \sim Z^{-\tau}$ \cite{Hirsh84},
as predicted by the classical Fisher droplet model for the
vicinity of the critical point \cite{Fisher67}. In \cite{Hirsh84} the critical temperature was
estimated to be $ \sim $ 5 MeV simply from the fact that the IMF mass
distribution is well described by a power law for the collision of $p$ (80-350
GeV) with Kr and Xe. In the paper \cite{Panagiotou85} the experimental data were gathered
for different colliding systems to get the temperature dependence of the
power law exponent. The temperature was derived from the inverse slope of
the fragment energy spectra in the range of the high-energy tail. The
minimal value of \textit{$\tau $} was obtained at $T$ = 11-12 MeV, which was claimed as
$T_{c}$. The later data smeared out this minimum. Moreover, it became clear
that the ``slope'' temperature for fragments does not coincide with the
thermodynamic one, which is significantly smaller. A more sophisticated use
of Fisher's model has been made in \cite{Elliott02}. The model is modified by including
the Coulomb energy release, when a particle moves from the liquid to the
vapor. The data for multi-fragmentation in \textit{$\pi $ }(8 GeV/c) + Au collisions were
analyzed. The extracted critical temperature was (6.7$\pm $0.2) MeV. The
same analysis technique was applied for collisions of Au, La, Kr (at 1.0 GeV
per nucleon) with a carbon target \cite{Elliott03}. The extracted values of $T_{c}$ are
(7.6$\pm $0.2), (7.8$\pm $0.2) and (8.1$\pm $0.2) MeV respectively.

Significantly higher critical temperature, 16.6$\pm $0.86 MeV, was obtained
in \cite{Natowitz02} by semi-empirical analysis of the data for the ``limiting
temperatures'' of fragmenting systems. Authors of Ref. \cite{Natowitz02} interpreted the
obtained value as $T_{c }$for the symmetric nuclear matter.

Having in mind the shortcomings of Fisher's model \cite{Schmelzer97, Reuter01} we have estimated
the nuclear critical temperature in the framework of the statistical
multi-fragmentation model, SMM \cite{Bondorf95}. This model describes well the different
properties of thermal disintegration of target spectators produced in
collisions of relativistic light ions. The intermediate mass fragment (IMF)
yield depends on the contribution of the surface free energy to the entropy
of a given final state. The surface tension coefficient of hot nuclei
depends on the critical temperature. The comparison of the measured and
calculated IMF charge yields provides a way to estimate $T_{c}$. It was found
from the analysis of the fragment charge distributions for the
$p$(8.1GeV)+Au reaction that $T_{c}$= (20 $\pm $ 3) MeV \cite{Karnaukhov03}. In the next paper
by the FASA collaboration \cite{Karnaukhov04} the value $T_{c}=(17\pm  2)$~ MeV was
obtained from an analysis the same data using a slightly different
separation of the events.

Thus, the different experimental estimations of the critical temperature
from fragmentation data are very controversial. This is a reason to look for
other observables which are sensitive to the critical temperature for the
liquid-gas phase transition. It was suggested in Ref. \cite{Karnaukhov97} to analyze the
temperature dependence of the fission probability to estimate $T_{c}$. Note,
that Silva et al. \cite{Silva04} explains why the power law used in the Fisher droplet
model gives a spurious value for $T_{c}$.

\section{TEMPERATURE DEPENDENCE OF FISSION BARRIER}

The fissility of heavy nuclei is determined by the ratio of the Coulomb and
surface free energies: the larger the ratio, the smaller the fission
barrier. As the temperature approaches the critical one from below, the
surface tension (and surface energy) gradually decreases, and the fission
barrier becomes lower. Thus, the measurement of fission probabilities for
different excitation energies allows an estimate of how far the system is
from the critical point. Temperature effects in the fission barrier have
been considered in a number of theoretical studies based on different models
(see e.g. \cite{Sauer76, Hasse73, Iljinov78, Pi82, Bartel85, Brack85, Garcias90}.
The effect is so large for hot nuclei that the barrier
vanishes, in fact, at temperatures of 4-6~MeV for critical temperature
$T_{c}$ in the range 15-18~MeV.

In terms of the standard liquid-drop conventions \cite{Nix68}, the fission barrier
can be represented as a function of temperature by the following relation:


\begin{eqnarray}
\label{eq1}
B_f(T,T_s)=E_s(T_s)-E_s^0(T)+E_c(T_s)-E_c^0(T)= \nonumber\\
E_s^0(T)\left[{(B_s-1)+ 2x(T)\cdot(B_c -1)}\right]
\end{eqnarray}

Here $B_{s}$ is the surface (free) energy at the saddle point in units of
surface energy$ E_{s}^{o}(T)$ of a spherical drop; $B_{c}$ is the Coulomb
energy at the saddle deformation in units of Coulomb energy
$E_{c}^{o}(T)$ of the spherical nucleus; $T_{s}$ and $T $are temperatures for the
saddle and ground state configurations. For the surface energy and the
fissility parameter $x(T)$, one can write \cite{Hasse73}:

\begin{eqnarray}
\label{eq2}
E_s^0 (T) = E_S^0 (0) \frac{\sigma (T)}{\sigma (0)} \left[
{\frac{\rho (0)}{\rho (T)}}\right]^{2 / 3},
\quad
x(T) = \frac{E_c^0 (T)}{2E_s^0 (T)} = x(0) \frac{\rho (T)\sigma
(0)}{\rho (0)\sigma (T)}
\end{eqnarray}

\noindent
where $\sigma (T)$ and $\rho (T)$ are the surface tension and the mean nuclear
density for a given temperature. Equation (\ref{eq1}) can be written as:

\begin{equation}
\label{eq3}
B_f (T,T_s) = B_f(T_s)+ \Delta B_f \\
\end{equation}

where $\Delta B_f=E_s^0(T_s )-E_s^0(T)+E_c^0(T_s)-E_c^0(T)$.
Here $B_{f }(T_{s})$ is fission barrier calculated under assumption that
$T_{s}=T$. In that case the values $B_{s}$ and $B_{c}$ are determined by the
deformation at the saddle point, which depends on the fissility parameter
$x(T). $These quantities were tabulated by Nix \cite{Nix68} for the full range of the
fissility parameter. The value of $\Delta B_{f}$ is determined by the
surface and Coulomb energies of a spherical drop, and can be easily
calculated. For $\sigma(T)$ we use the approximation:

\begin{equation}
\label{eq4}
\sigma(T)=\sigma(0)\left[ {\frac{T_c^2-T^2}{T_c^2+T^2}}\right]^{5/4}
\end{equation}

This equation was obtained in Ref. \cite{Ravenhall83} devoted to the consideration of
thermodynamic properties of a plane interface between liquid and gaseous
phases of nuclear matter in equilibrium. This parameterization is
successfully used by the SMM for describing the multi-fragment decay of hot
nuclei. Figure~1 shows the different approximations used in the literature
for the surface tension coefficient as a function of $T$/$T_{c}$.

\begin{figure}[ht]
\includegraphics[width=0.7\textwidth]{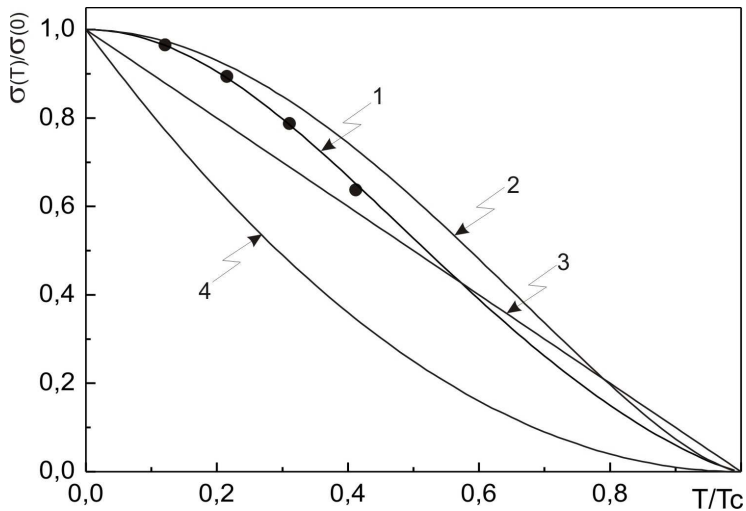}
\caption{The calculated coefficient of the surface tension as a
function of $T/T_{c}$: lines 1 and 2 are obtained according to eq. (\ref{eq4}) and
\ref{eq5}), lines 3, 4 are for linear and quadratic parameterizations of $\sigma(T)$.
The symbols are taken from Ref.[1].}
\label{fig1}
\end{figure}

Curve number 2 was calculated in framework of semi-classical model, based on
the Seyler-Blanchard interaction \cite{Lima92}. An analytical expression for $\sigma
(T)$ obtained in this paper is the following:

\begin{equation}
\label{eq5}
\sigma (T)=\sigma(0)\left( {1 + 1.5\frac{T}{T_c}} \right)
\left({1-\frac{T}{T_c}}\right)^{1.5}
\end{equation}

Two other parameterization of $\sigma (T)$ are also presented: linear $\sim $
(1-$T$/$T_{c})$, which is used in the analysis with the Fisher droplet model
\cite{Elliott02, Elliott03}, and quadratic $\sim $ (1-$T$/$T_{c})^{2}$ \cite{Richert01}.

In accordance with \cite{Nix68}, the expressions for $E_s^0 (0)$ and $x$(0) are taken to
be

\begin{equation}
\label{eq6}
E_s^0 (0) = 17,9439 \gamma  \cdot A^{2/3} MeV, \,\,\
x(0)=\frac{{Z^2 /A}}{{50.88\gamma }},
\end{equation}

where  $\gamma=1-1.7826\left[{\frac{{N - Z}}{A}}\right]^2$.
Sauer \textit{et al. } \cite{Sauer76} investigated the thermal properties of nuclei by using the
Hartree-Fock approximation with the Skyrme force. The equation of state was
obtained, which gives the critical temperature $T_{c}\approx $ 18 MeV for
finite nuclei. The temperature dependence of the mean nuclear density was
found to be \textit{$\rho $}($T)$\textit{ =$\rho $}(0)(l - $\alpha $\textsc{\textit{T}}$^{2}$\textsc{), }where
$\alpha=1.26 \cdot 10^{-3}$ MeV$^{ - 2}$. In the following we
shall use this finding for\textit{ $\rho $}($T)$.

\begin{figure}[ht]
\includegraphics[width=0.7\textwidth]{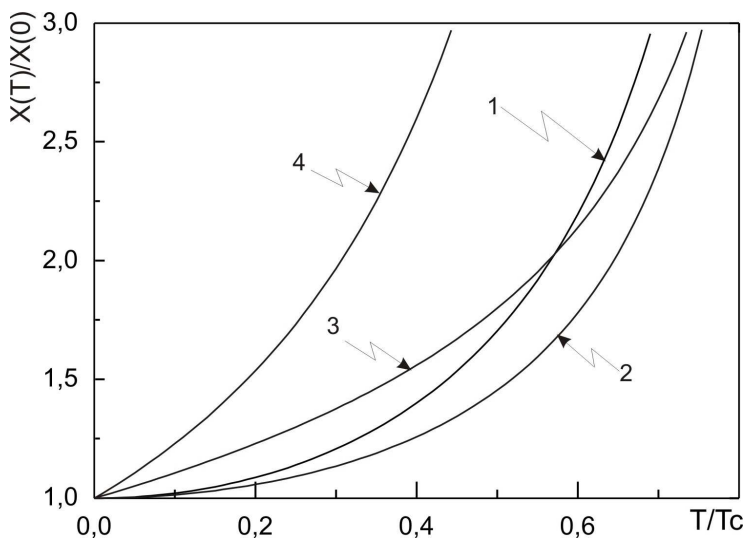}
\caption{Relative value of fissility parameter, calculated for
$^{l88}$Os as a function of relative temperature for different parameterization
of surface tension. Meaning of the lines is explained in caption of Fig.1.}
\label{fig2}
\end{figure}

Figure 2 shows the relative values of the fissility parameter $x(T)$ for
$^{l88}$Os calculated as a function of reduced temperature$ T$/$T_{c}$. This
nucleus has been chosen since the results can be compared with well known
experimental data for this nucleus \cite{Moretto72}. The calculations are performed for
the different versions of $\sigma(T)$ mentioned above. A drastic change of
nuclear fissility is expected even halfway to the critical point.

Figure 3 displays the calculated value of the liquid-drop fission barrier
for $^{188}$Os as a function of relative temperature. Virtually, the barrier
vanishes for $T>0.4T_{c}$ if the surface tension is taken according to (\ref{eq4})
and (\ref{eq5}). For the linear and quadratic approximations of $\sigma $ ($T)$ the
reduction of the fission barrier with temperature is much faster.

\begin{figure}[ht]
\includegraphics[width=0.7\textwidth]{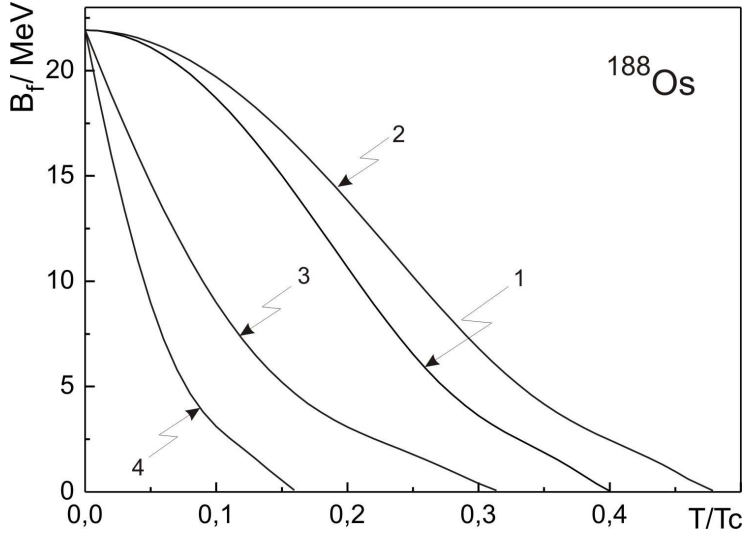}
\caption{Temperature dependence of liquid - drop fission barrier for
$^{188}$Os. The meaning of the lines is explained in Fig.1 caption.}
\label{fig3}
\end{figure}

\section{THE ESTIMATION OF FISSION PROBABILITY}

In this chapter we analyze the experimental data on the fission probability
of $^{188}$Os, produced in collisions $^{4}He+^{184}$W \cite{Moretto72}. The
excitation energy of the compound nucleus created at the highest beam energy
is 117 MeV. The shell and pairing effects are predicted to disappear for
such a hot nucleus; therefore the fission barrier is expected to be the
liquid-drop one. This barrier is temperature dependent. Comparison of the
measured and model calculated fission probabilities provides a way to
estimate the critical temperature $T_{c}$.

Experimentally the fission probability $W_{f}$ can be found from the measured
fission cross section $\sigma_{f}$:

\begin{equation}
\label{eq7}
W_f = \sigma _f / \sigma _R ,
\end{equation}

\noindent
where $\sigma_{R}$ is total reaction cross-section. The main decay mode of the
compound nucleus in $^{4}He +^{184}$W collisions is the sequential
emission of neutrons. For the highest excitation energy the mean number of
emitted neutrons is 11-12. The mean fission probability during a neutron
cascade of$x$ steps can be calculated by the following equation:

\begin{equation}
\label{eq8}
W_f =1-\prod\limits_{i = 1}^x {\left[{1-\frac{\Gamma_f(A_i,Z_i,E_i^\ast
)}{\Gamma_{tot}(A_i,Z_i,E_i^\ast )}}\right]} ,
\end{equation}

The ratio\textit{ $\Gamma $}$_{f }$/\textit{$\Gamma $}$_{tot }$is the relative fission width for the $i$-step of the
cascade. According to the statistical model \cite{Bass80} the value of\textit{ $\Gamma $}$_{f }$is
calculated as

\begin{equation}
\label{eq9}
\Gamma_f (E_i^\ast ,I_i ) = \frac{1}{2\pi \cdot \rho (U_i )}\int\limits_0^{U_i -
B_{fi}} {\rho_S(U_i-B_{fi} - \varepsilon )d\varepsilon }
\end{equation}

Here $U$ is the thermal part of excitation energy $E$*, \textit{$\rho $}($U)$ is the level density,
the index $s$ is used for the saddle configuration. It is natural to use in (\ref{eq3})
the temperature dependent fission barrier as has been done
in \cite{Hasse73, Iljinov78,Pi82, Bartel85, Brack85, Garcias90}.
The problem was considered also in \cite{Charity96}. The neutron width is given by the
following equation \cite{Barashenkov72}:

\begin{equation}
\label{eq10}
\Gamma_n (E_i^\ast ,I_i ) = \frac{2(2S_n + 1)m_n }{\pi ^2h^3\rho_i(U_i
)}\int\limits_0^{U_i-B_{ni} } {\sigma_n(E_n)\rho_i(U_i-B_{ni}}-
E_n )E_n dE_n
\end{equation}

Here $B_{ni}, E_{n}, S_{n}$ are binding, kinetic energies and spin of the
neutron, $\sigma_{n}(E_{n})$ is the neutron capture cross-section for the inverse
reaction. The contribution of charged particle evaporation is on the level
of several percent of $\Gamma_{total}$. Nevertheless it has been taken into
account. For level density $\rho(U)$ the Fermi-gas model is used.

Figure 4 presents the comparison of the data for fissility of $^{188}$Os as
a function of excitation energy \cite{Moretto72} with calculations under the assumption
that the surface tension is described by eq. (4). The critical temperature
is a parameter that can be found from the best fit. It is done for the
highest excitation energy available, where the temperature dependence of the
fission barrier is more prominent.

\begin{figure}[ht]
\includegraphics[width=0.7\textwidth]{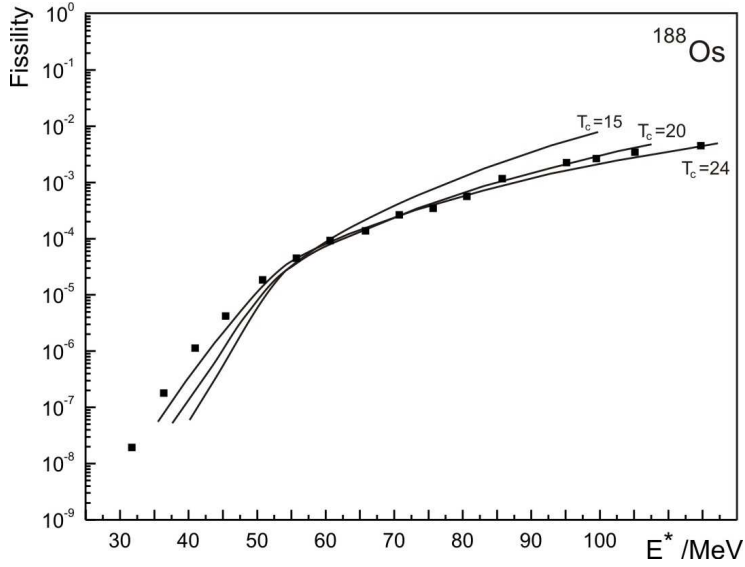}
\caption{Fission probability of $^{188}$Os as a function of the
excitation energy: dots are data \cite{Moretto72}, curves are calculated assuming
different values of critical temperature. Surface tension is taken according
to (4).}
\label{fig4}
\end{figure}

The result is demonstrated in Fig.5. Different calculations are presented,
which have been done using all the parameterization of the surface tension
mentioned above. It seems clear that the linear and quadratic approximations
for $\sigma(T)$ should be excluded as unrealistic. Fission probabilities,
calculated with egs. (4) and (5) fall rather fast with increasing the
critical temperature. They are crossing the experimental band giving the
following values of critical temperature: $T_{c}\approx $ (23.5 $\pm $
2.5) MeV in the first case, and at $T_{c } \approx $ (17.5 $\pm $ 1.5) MeV
for the use of eq.(5). This is in accordance with the value of the critical
temperature obtained by the FASA collaboration from multi-fragmentation
data. These values are only slightly changed when the shell effect is taken
into account for the last steps of the neutron cascade.

\begin{figure}[ht]
\includegraphics[width=0.7\textwidth]{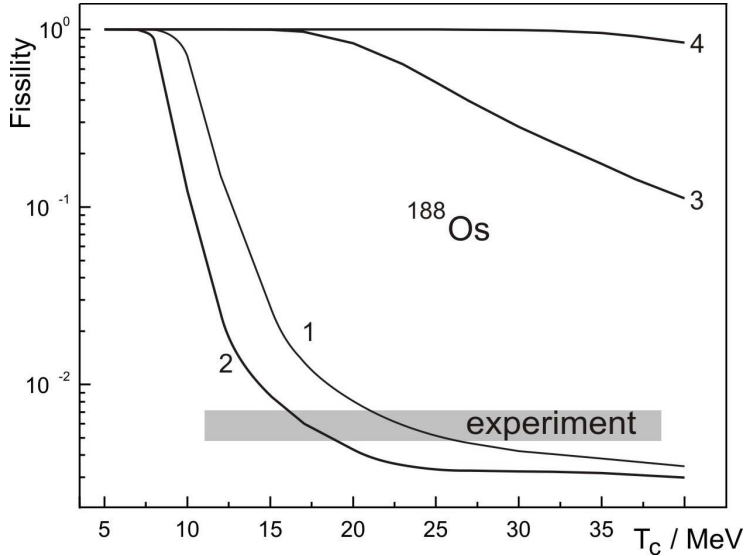}
\caption{Fission probabilities for $^{188}$Os at excitation energy
117~MeV. The calculated values (lines) are given as a function of the
assumed critical temperature. Different parameterizations of surface tension
are used (see Fig.1). The experimental value is shown by the horizontal
band.}
\label{fig5}
\end{figure}

\section{CONCLUSION}

Critical temperature for the nuclear liquid-gas phase transition has been
estimated from the fission probability of the highly excited nucleus
$^{188}$Os. Analysis is made under different assumptions about the
temperature dependence of nuclear surface tension. The results presented
here provide strong support for the value $T_{c}\ge $ 16 MeV.

Authors are indebted to E.Norbeck and H. Oeschler for illuminating discussions. The
research was supported in part by the Russian Foundation for Basic Research,
Grant 06-02-16068 and the Grant of Polish Plenipotentiary to JINR.


\begin{thebibliography}{99}


\bibitem{Sauer76} G. Sauer, G. Chandra H. and U. Mosel, Nucl. Phys. A264 (1976) 221.
\bibitem{Jaqaman83} H. Jaqaman, A.Z. Mekjian and A.Z. Zamick L., Phys. Rev. C 27 (1983) 2782.
\bibitem{Shou96} Zhang Feng Shou, Z. Phys. A 356 (1996) 163.
\bibitem{Goodman84} A.L. Goodman, J.I. Kapusta and A.Z. Mekjian, Phys. Rev. C 30 (1984) 851.
\bibitem{Taras04} S. Taras et al., Phys. Rev. C 69 (2004) 014602.
\bibitem{Randrup91} J. Randrup and E. de Lima Medeiros, Nucl. Phys. A529 (1991) 115.
\bibitem{Lima92} E. de Lima Medeiros and J. Randrup, Phys. Rev. C 45 (1992) 372.
\bibitem{Hirsh84} A.S. Hirsh et al., Phys. Rev. C 29 (1984) 508.
\bibitem{Fisher67} M.E. Fisher, Physics (N.Y.) 3 (1967) 255.
\bibitem{Panagiotou85} A.D. Panagiotou et al., Phys. Rev. C 31 (1985) 55.
\bibitem{Elliott02} J.B. Elliott et al., Phys. Rev. Lett. 88 (2002) 042701.
\bibitem{Elliott03} J.B. Elliott et al., Phys. Rev. C 67 (2003) 024609.
\bibitem{Natowitz02} J. Natowitz et al., Phys. Rev. Lett. 89 (2002) 212701.
\bibitem{Schmelzer97} J. Schmelzer, G. R\"{o}pke, F.P. Ludwig, Phys. Rev. C 55 (1997) 1917.
\bibitem{Reuter01} P.T. Reuter, K.A. Bugaev, Phys. Lett. B 517 (2001) 233.
\bibitem{Bondorf95} J.P. Bondorf et al., Phys. Rep. 257, ¹3 (1995) 134.
\bibitem{Karnaukhov03} V.A. Karnaukhov et al., Phys. Rev. C 67 (2003) 011601(R).
\bibitem{Karnaukhov04} V.A. Karnaukhov et al., Nucl. Phys. A734 (2004) 520.
\bibitem{Karnaukhov97} V.A. Karnaukhov, Phys. At. Nucl. 60 (1997) 1625.
\bibitem{Silva04} J.D. Silva et al., Phys. Rev. C 69 (2004) 024606.
\bibitem{Hasse73} R.W. Hasse and W. Stocker, Phys. Lett. B 44 (1973) 26.
\bibitem{Iljinov78} A.S. Iljinov, E.A.Cherepanov, S.E.Chigrinov, Z. Phys. A 287 (1978) 37.
\bibitem{Pi82} M. Pi et al., Phys. Rev. C 26 (1982) 773.
\bibitem{Bartel85} J. Bartel, P. Quentin, Phys. Lett. B 152 (1985) 29.
\bibitem{Brack85} M. Brack et al., Phys. Rep. 123 (1985) 275.
\bibitem{Garcias90} F. Garcias et al., Z. Phys. A: At. Nucl. 336 (1990) 31.
\bibitem{Nix68} J. Nix, Nucl. Phys. A. 13 (1968) 241.
\bibitem{Ravenhall83} D.G. Ravenhall et al., Nucl. Phys. A40 (1983) 71.
\bibitem{Richert01} J. Richert and P. Wagner, Phys. Rep. 350 (2001) 1.
\bibitem{Moretto72} L.G. Moretto et al., Phys. Lett. B 38 (1972) 471.
\bibitem{Bass80} R. Bass, Nuclear Reactions with Heavy Ions, Springer-Verlag, 1980.
\bibitem{Charity96} R.J. Charity, Phys. Rev. C 53 (1996) 512.
\bibitem{Barashenkov72} V.S.Barashenkov, V.D.Toneev: Interaction of high energy particles and nuclei
with nuclei, p.401, Moscow: Atomizdat 1972.

\end{thebibliography}
\end{document}